# GC$_3$ Biology in Eukaryotes and Prokaryotes

Eran Elhaik[1,2] and Tatiana Tatarinova[3]
*[1]McKusick - Nathans Institute of Genetic Medicine,
Johns Hopkins University School of Medicine, Baltimore,
[2]Department of Mental Health,
Johns Hopkins University Bloomberg School of Public Health,
[3]Faculty of Advanced Technology, University of Glamorgan*
[1,2]*USA*
[3]*Wales*

## 1. Introduction

In this chapter we describe the distribution of Guanine and Cytosine (GC) content in the third codon position (GC$_3$) distributions in different species, analyze evolutionary trends and discuss differences between genes and organisms with distinct GC$_3$ levels. We scrutinize previously published theoretical frameworks and construct a unified view of GC$_3$ biology in eukaryotes and prokaryotes.

## 2. The wobble position

### 2.1 Why is GC$_3$ referred to as the wobble position

The genetic code, the set of rules by which information encoded in genetic material, is found in every cell of every living organism. This code consists of all possible combinations of tri-nucleotide sequences in coding regions, called codons. With a few exceptions, such as start- and stop-codons, a triplet codon in a DNA sequence specifies a single amino acid – protein's building block. The human genome, for example, consists of one start- and three stop-codons out of ($3^4$) 64 codons. With each codon promoting the binding of specific tRNA to the ribosome, the cell would theoretically need almost 64 types of tRNAs, each with different anticodons to complement the available codons. However, because only 20 amino acids are encoded, there is a significant degeneracy of the genetic code so that the third base is less discriminatory for the amino acid than the other two bases. This third position in the codon is therefore referred to as *the wobble position*. At this position U's and C's may be read by a G in the anticodon. Similarly, A's and G's may be read by a U or ψ (pseudouridine) in the anticodon.

### 2.2 Identification of the mRNA codons by tRNA

For most amino acids, there are specific enzymes ligating their cognate amino acid to the tRNA molecule-bearing anticodon that correspond to that amino acid. These enzymes and the unique structure of each tRNA ensure that a particular tRNA is the substrate for its cognate synthetase and not for all other syntheses present in the cell (Shaul *et al.* 2010). Like in the genetic code, there is much redundancy in the types of tRNA molecules required per cell. Because the wobble base positions are capable of binding to several codons, a minimum set of 31 tRNA are required to unambiguously translate all the codons instead of 61 types tRNA molecule required to match each codon. This redundancy in tRNA anticodons is accomplished, in part, by using inosine, which can pair with U, C, or A, at the third position of the mRNA (Ikemura 1985).

## 3. Biological role of $GC_3$: Codon usage, codon bias, mRNA, gene expression, gene and promoter organization, gene function, and methylation

Deviations from unimodal bell-shaped distributions of $GC_3$ appear in many species (Aota and Ikemura 1986; Belle, Smith, and Eyre-Walker 2002; Jorgensen, Schierup, and Clark 2007). This bimodality in homeotherm (termed, "warm-blooded", at that time) vertebrates was originally explained by the presence of *isochores* – long (>300,000 bp) and relatively homogeneous stretches of DNA (Mouchiroud, Fichant, and Bernardi 1987). Although there are similarities between genes in high-GC human isochores and $GC_3$-rich genes in grasses, the isochore hypothesis does not fully explain the existence of $GC_3$-rich genes in grasses: first, there is no correlation between GC contents of open reading frames (ORFs) and the flanking regions; second, most species with isochores do not have a $GC_3$-rich peak (Tatarinova *et al.* 2010). Therefore, the remaining possible causes of bimodality may be elucidated by comparing genes in different $GC_3$ classes, such as $GC_3$-rich and $GC_3$-poor classes. These classes differ in nucleotide composition and compositional gradient along coding regions (Figure 1). $GC_3$-rich class genes have a significantly higher frequency of CG dinucleotides (potential targets for methylation); therefore, there is an additional regulatory mechanism for $GC_3$-rich genes. Springer *et al.* 2005 reported that out of eight classes of methyl-CpG-binding domain proteins present in dicots, only six exist in monocots, suggesting a difference between dicots and monocots in silencing of methylated genes.

In 2010 Tatarinova, Alexandrov, Bouck and Feldmann proposed the following explanation of relationship between DNA methylation and $GC_3$ content. Two competing processes may affect the frequency of methylation targets: the GC-based mismatch repair mechanism and AT-biased mutational pressure. In recombining organisms (e.g., grasses and homeotherms vertebrates), the GC content of coding and regulatory regions is enhanced because of the action of the GC-based mismatch repair mechanism; this effect is especially pronounced for $GC_3$. Recombination has been shown to be a driving force for the increase in $GC_3$ in many organisms. Repair (recombination) happens all over the genome with a certain precision, leading to an increase in GC content. If repair did not occur in defense-related genes, the organism may fail to survive or to reproduce. However, if repair did not happen in genes that are not essential for survival, and, consequently, their GC content remained the same, it may not be detrimental to the organism. AT-biased mutational pressure, resulting from cytosine deamination or oxidative damage to C and G bases, counteracts the influence of recombination; and in most asexually-reproducing species and self-pollinating plants, AT bias is the winning process. Our analysis from aligning *Indica* and *Japonica* indicates that genomic regions under higher selective pressure are more frequently recombining and therefore increasing their $GC_3$ content (Tatarinova *et al.* 2010). This mechanism may explain the pronounced differences in $GC_3$ between *A. thaliana* and its closest relatives. Comparison of the nucleotide compositions of coding regions in *A. thaliana, R. sativus, B. rapa*, and *B. napus* reveals that the $GC_3$ values of *R. sativus, B. rapa*, and *B. napus* genes are on average 0.05 higher than those of the corresponding *A. thaliana* orthologs.

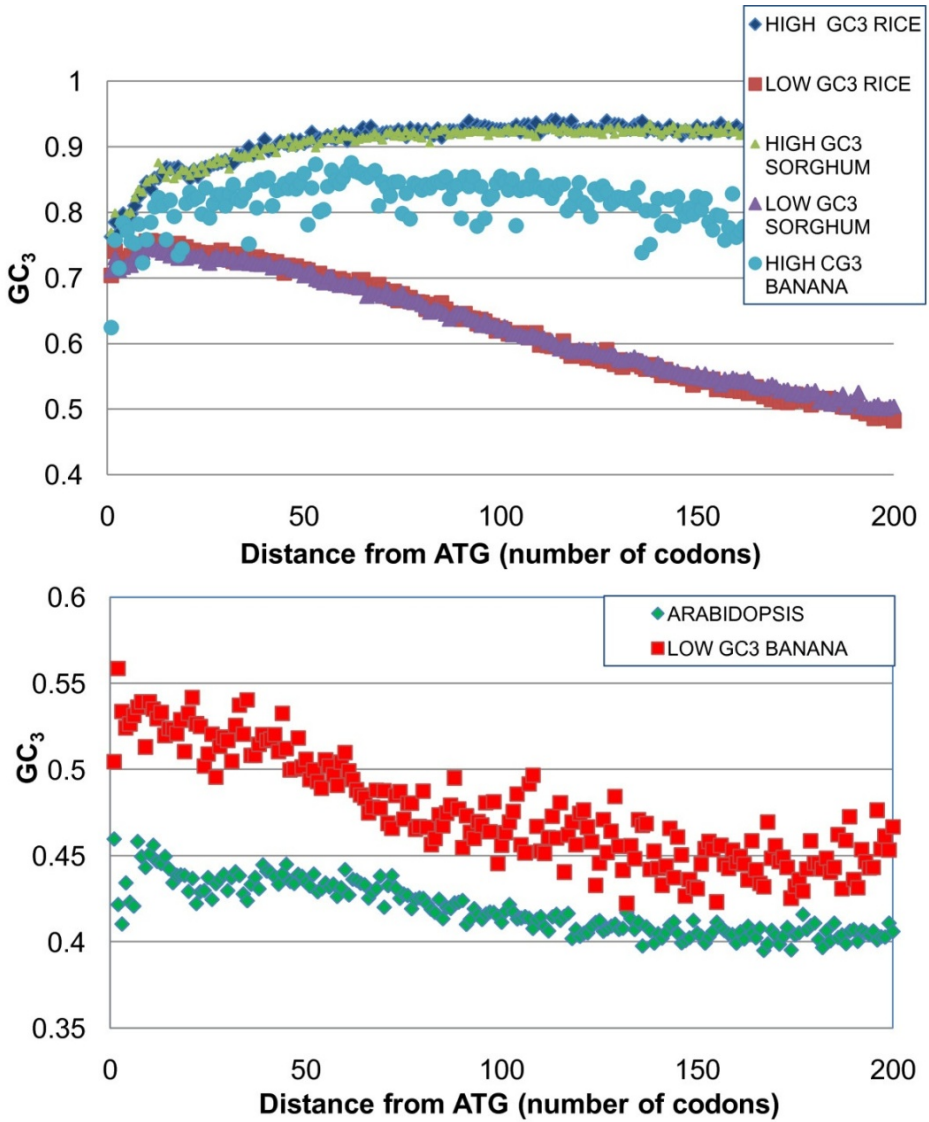

Fig. 1. $GC_3$ gradient from 5' to 3' ends of coding regions. At the 5' end of the open reading frame, high $GC_3$ genes of rice, sorghum, and banana have a slight positive gradient, whereas low $GC_3$ genes in arabidopsis, rice, sorghum, and banana become more $AT_3$-rich (From Tatarinova *et al* 2010)

An important difference between *A. thaliana* and *Brassica* and *Raphanus* is that the latter two genera are self-incompatible, whereas *A. thaliana* is self-pollinating. Self-pollination in arabidopsis keeps its recombination rates low and thus reduces the $GC_3$ content of its genes. Self-pollination is also reported in some grasses, such as wheat, barley and oats. Analysis of recombination in wheat showed that the genome contains areas of high and low recombination. Grasses have an efficient reproductive mechanism and high genetic variability that enables them to adapt to different climates and soil types. We hypothesize that since self-pollination generally lowers recombination rates, evolutionary pressure will selectively maintain high recombination rates for some genes. Analysis of highly recombinogenic genomic regions of wheat, barley, maize, and oat identified several genes of agronomic importance in these regions (including resistance genes against obligate biotrophs and genes encoding seed storage proteins) (Keller and Feuillet 2000). In addition to the methylation-driven growth of high-$GC_3$ genes, we hypothesize that the development of $GC_3$-richness in some genes may, if unbalanced by AT-bias, work as a feed-forward mechanism. Once $GC_3$-richness appears in genes under selective pressure, it provides additional transcriptional advantage. GC pairs differ from AT pairs in that guanine binds to cytosine with three hydrogen bonds, while adenine forms only two bonds with thymine. This additional hydrogen bond makes GC pairs more stable; thus GC-rich genes will have different biochemical properties from AT-rich genes. When an AT pair is replaced by a GC pair in the third position of a codon, the protein sequence remains largely unchanged, but an additional hydrogen bond is introduced. This additional bond can make transcription more efficient and reliable, change the array of RNA binding proteins or significantly alter the three-dimensional folding of the messenger RNA. In this case, those plant species that thrive and adapt successfully to harsh environments demonstrate a strong preference for GC in the third position of the codon.

High $GC_3$ content provides more targets for methylation. The correlation between methylation and $GC_3$ is supported by Stayssman *et al*. (2009), who reported a positive correlation between methylation of internal unmethylated regions and expression of the host gene. In this paper the authors have demonstrated a positive correlation between $GC_3$ and variability of gene expression; they also found that $GC_3$-rich genes are more enriched in CG than the low-$GC_3$-poor gene class. Therefore, $GC_3$-rich genes provide more targets for *de novo* methylation, which can serve as an additional mechanism of transcriptional regulation and affect the variability of gene expression. Overall, additional transcriptional regulation makes species more adaptable to external stresses.

## 4. Genome-wide view

### 4.1 GC$_3$ in animals

The $GC_3$ varies substantially within animal genomes. Animals can be divided to homeotherms-those that maintain a stable internal body temperature, like mammals and birds, and sometimes termed "warm-blooded" and poikilotherms - those whose internal temperature varies considerably and are often termed "cold-blooded." These differences were integrated into molecular evolution by Mouchiroud *et al*. (1987) who argued that in poikilotherm vertebrate, genes are mostly GC-poor and are harbored by GC-poor intergenic regions, whereas most genes of homeothermy vertebrates are GC-rich and found predominantly on the scant GC-rich intergenic regions. Because $GC_3$ was shown to be correlated to the GC content of the gene, it became the primary tool to study the differences between homeotherms and poikilotherms. Indeed, poikilotherms were shown to have lower $GC_3$ than homeotherms on average, although some poikilotherms exhibit higher $GC_3$ values than homeotherms. Most poikilotherms also exhibit a lower variation in $GC_3$ and a correlation to the GC content of the first two codon positions, indicating a systemic compositional variation across their genome (Belle, Smith, and Eyre-Walker 2002).

$GC_3$ exhibits a wide variation in different animals compared with the average GC content (Figure 2). In humans, $GC_3$ ranges from 22 to 97% compared with the range of GC content (32-80%), and in zebrafish the $GC_3$ range is more limited 27-92% ($\mu$=56%, $\sigma$=8%) yet still wider than the GC content range (34-68%) (Figure 3) (Elhaik, Landan, and Graur 2009). Because $GC_3$ is mostly unconstrained by functional requirements, that is, by the need to code specific amino acids and because $GC_3$ exhibits a non-uniform distribution, the third-codon position became a natural candidate to investigate the forces that shaped the composition of the genome.

| Species | No. of genes | $GC_3$ | | | GC content | | |
|---|---|---|---|---|---|---|---|
| | | Mean | σ | Range | Mean | σ | Range |
| *Homo sapiens* | 17,451 | 0.6 | 0.17 | 0.22-0.97 | 0.45 | 0.06 | 0.32-0.8 |
| *Bos Taurus* | 5,522 | 0.62 | 0.16 | 0.25-0.97 | 0.43 | 0.06 | 0.33-0.76 |
| *Mus musculus* | 17,009 | 0.59 | 0.11 | 0.21-0.96 | 0.43 | 0.05 | 0.27-0.76 |
| *Rattus norvegicus* | 8,983 | 0.59 | 0.11 | 0.23-0.96 | 0.42 | 0.06 | 0.33-0.73 |
| *Gallus gallus* | 3,036 | 0.56 | 0.15 | 0.28-0.99 | 0.42 | 0.05 | 0.36-0.8 |
| *Danio rerio* | 4,344 | 0.56 | 0.08 | 0.27-0.92 | 0.35 | 0.02 | 0.34-0.68 |

Fig. 2. $GC_3$ and GC content for 6 Vertebrate Taxa. From (Elhaik, Landan, and Graur 2009).

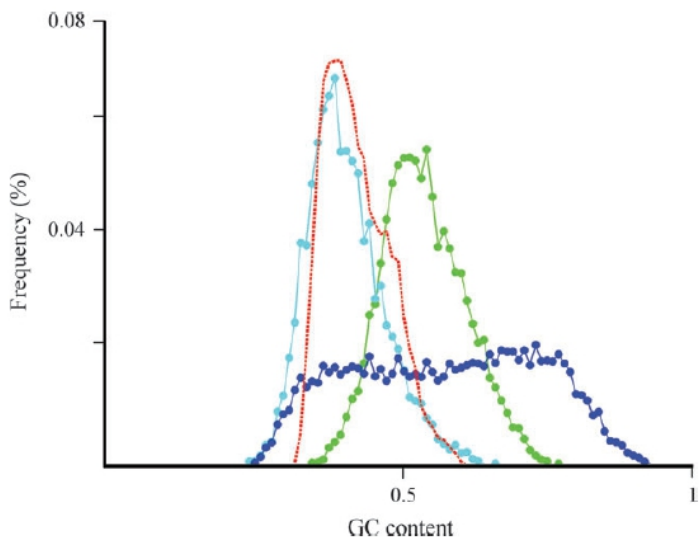

Fig. 3. GC content in codon positions: $GC_1$ (green), $GC_2$ (turquoise), $GC_3$ (blue), and 200-kb flanking regions (dashed red) in human. From (Elhaik, Landan, and Graur 2009).

### 4.2 GC$_3$ in yeast

The variation in GC content along yeast chromosomes was first reported in *S. cerevisiae* (Sharp and Lloyd 1993) where GC$_3$ ranged from 35% to 50% on chromosome III. Later observations confirmed similar patterns for the remaining chromosomes and linked the GC variation to recombination, that is, rich GC$_3$ regions would have a higher recombination rate and recombination hot spots would occur in high picks of GC content (Bradnam *et al*. 1999). Findings from eukaryotic genomes suggested that biased gene conversion (see below) may be the molecular mechanism that facilitated the increase of GC$_3$ in recombination hot spots.

The large variation in GC$_3$ also exists between yeast species with *C. tropicalis* having the lowest median GC$_3$ (22%) and *C. lusitaniae* with a GC$_3$ median of 49%. Overall, species that were more closely related to each other tended to have more similar GC$_3$ distributions (Figure 4). Despite of these vast differences, the locations of peaks and troughs of GC$_3$ largely coincide. In other words, the differences in base composition between the species varies systematically across all genes, with higher divergence in GC-rich genes than GC-poor genes (Lynch *et al*. 2010). The conserved nature of the GC$_3$ variation along chromosomes and the coinciding peaks and troughs led Lynch *et al*. (2010) to propose that the GC-poor troughs indicate the positions of ancient centromeres at the points where deep GC-poor regions were found.

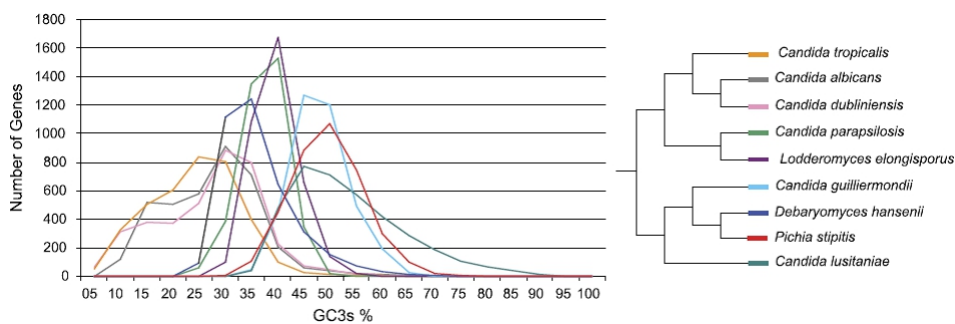

Fig. 4. Distribution of GC$_{3s}$ values among nine species in the Candida clade. A set of 3,687 orthologous genes was identified among all these species. The phylogenetic tree on the right is modified from Butler *et al*. (2009) and is derived from 644 single-gene families. From (Lynch *et al*. 2010).

### 4.3 GC$_3$ in plants: Adaptation to environment and codon usage (Tatarinova et al. 2010)

Pronounced differences in GC$_3$ exist both within and between plant genomes. For example, GC$_3$ in rice genes ranges from 43% to 92% (Wang, Singer, and Hickey 2004). Grasses have undergone several genome duplications. Genomic regions varied in their recombination rates and GC$_3$ contents. Since high GC$_3$ content in a gene provided an evolutionary advantage, this was frequently the sole copy retained in grasses. This may explain why GC$_3$-rich genes frequently lack paralogs. GC$_3$-rich genes provide an evolutionary advantage because of their optimized codon usage and the existence of methylation targets allowing for an additional mechanism of transcriptional regulation. Therefore, GC$_3$-rich genes were maintained in grasses for generations.

The evolutionary forces affecting development of plants are realized through introducing new mutations during meiotic recombination and fixation with the help of DNA methylation and transcriptional mechanisms. The presence of $GC_3$-rich genes is not likely to be a consequence of chromosomal isochores or horizontal gene transfer. Regardless of their initial origin, $GC_3$-rich genes in recombining species possessed a self-maintaining mechanism that over time could only increase their drift towards even higher $GC_3$ values. This uncompensated drift may explain the pronounced bimodality of some rapidly-evolving species. Competing forces acting in grasses make $GC_3$ distribution distinctly bimodal; $GC_3$-rich genes are more transcriptionally regulated, provide more targets for methylation and accumulate more mutations than $GC_3$-poor genes.

### 4.4 $GC_3$ in prokaryotes

In prokaryotic genomes the nucleotide composition varies from the extremely low GC content (15%) in obligatory intracellular bacteria to the high GC content (75%) in Proteobacteria and from 27% to 66% in Archaea. (http://www.ncbi.nlm.nih.gov/genomes/lproks.cgi). The base compositional deviations show tremendous variation even at the nucleotide level of the three codon positions. With $GC_1$ follows a global tendency of monotonic decrease versus the increase of the genomic GC content and the $GC_2$ follows a global monotonic increase, as expected. The $GC_3$ positions range from 10% to 90% (Muto and Osawa 1987) and exhibit a more complicated pattern that decreases first and increases last with a global minimum at about 40% of the genomic GC content (Ma and Chen 2005).

In many organisms, alternative synonymous codons are not used with equal frequency, hence the codon usage is considered biased. This bias exists not only between different organisms, but often among genes within a genome (Suzuki, Saito, and Tomita 2009). Different factors have been proposed to contribute to synonymous codon usage bias, including replication strand bias, translational selection, and GC composition (Ermolaeva 2001). Because codon third positions are largely degenerate - 70% of changes at third codon positions are synonymous and they are commonly considered correlated with synonymous codon usage bias, although in practicality, the strength of this correlation varies widely among species (Suzuki, Saito, and Tomita 2007). The large deviations in base composition of these sites were also thought to reveal the underlying mutational bias of the genome and served as the basis for the original formulation of the neutral theory (Sueoka 1962; Sueoka 1988).

### 4.5 The isochore theory

Prior to the publication of the draft human genome (Lander *et al*. 2001), scientists were limited to the study of genes and short (<500 bp) flanking regions. The publication of the draft human genome (Lander *et al*. 2001) in 2001 was quickly followed by the publications of fully sequenced genomes from other species (e.g., Chimpanzee, mouse, and cow), which enabled us to study the evolution of genomes whole biological entities, rather than as a collection of genes. One of the most common ways to describe a genome is by means of the nucleotide distribution, particularly the distribution of GC content. In the absence of genomic data, inferences made on short fragments were based on the assumption that these fragments represent the compositional complexity of the entire genome (e.g., Aissani *et al*. 1991; Mouchiroud and Bernardi 1993). The GC content patterns emerging from these analyses were used by Bernardi and colleagues (Macaya, Thiery, and Bernardi 1976; Thiery, Macaya, and Bernardi 1976; Bernardi *et al*. 1985) to explain the differences between the genome organization of "warm-blooded" and "cold-blooded" vertebrates (Cuny *et al*. 1981; Bernardi *et al*. 1985; Bernardi 2000) with the first described as a mosaic of GC-poor and GC-rich isochores and the later as devoid of GC-rich isochores (Bernardi *et al*. 1985).

Because $GC_3$ is mostly unconstrained by functional requirements, that is, by the need to code specific amino acids, the third-codon position was a natural candidate for a predictive proxy of flanking GC content. In spite of the lack of correlation between $GC_3$ and large flanking regions of "isochoric regions" harboring the genes (Bernardi 1993a), over time it became a common belief that such a relationship exists. Over the next two decades $GC_3$ was used extensively as the primary means to predict isochore structure, surprisingly enough, even after full genome sequences were made available. Many of the theories concerning the evolution of isochores are also based on studies that used $GC_3$ as a predictor for isochore composition or that simply assumed the existence of isochores. (Bernardi 2001; Ponger, Duret, and Mouchiroud 2001; D'Onofrio 2002; D'Onofrio, Ghosh, and Bernardi 2002; Hamada *et al*. 2003; Romero *et al*. 2003; Federico *et al*. 2004; Chojnowski *et al*. 2007; Fortes *et al*. 2007)

Two opposite explanations were proposed to explain the formation of isochores. The first view was that isochores may simply reflect variable mutation processes among genomic regions, consistent with the neutral model (Wolfe, Sharp, and Li 1989; Sueoka and Kawanishi 2000; Galtier *et al*. 2001). Alternatively, isochores were posited as results of natural selection for certain compositional environment required by certain genes (Matassi, Sharp, and Gautier 1999). It should be noted that these hypotheses are not mutually exclusive; two or more of the processes could be acting together (Eyre-Walker and Hurst 2001). For example, the most accepted hypothesis for the unequal usage of synonymous codons in bacterial genomes is that the unequal usage is the result of a very complex balance among different evolutionary forces (mutation and selection) (Suzuki, Saito, and Tomita 2007). Several hypothesis derive from the selectionist view, such as the biased gene conversion hypothesis (Galtier *et al*. 2001), the thermodynamic stability hypothesis (Bernardi and Bernardi 1986; Bernardi 1993b), the transposable elements hypothesis (Duret and Hurst 2001), the recombination hypothesis (Eyre-Walker 1993) and the cytosine deamination hypothesis (Fryxell and Zuckerkandl 2000).

The presumed relationship between $GC_3$ and isochores has been used numerous times in the literature to study isochore function and evolution until Elhaik *et al*. (2009) showed that no such relationship exists. By testing the relationship between $GC_3$ and the GC content of the flanking regions of the genes of 6 animals, the authors demonstrated that $GC_3$ explains a very small proportion of the variation in GC content of long genomic sequences flanking the genes. The predictive power either decreases rapidly the further one gets from the gene or does not exist at all. These findings also implied that the isochore theory cannot be discussed without further analysis of the complete genomic data. Indeed, further analyses showed that the descriptions of the human or vertebrate genomes as mosaics of isochores are erroneous (Cohen *et al*. 2005; Elsik *et al*. 2009; Elhaik, Graur, and Josić 2010; Elhaik *et al*. 2010). Due to the lack of predictive power of $GC_3$, new genomic studies scan the entire genomic structure using automatic algorithms rather than rely on unreliable proxies. The emerging view of the mammalian genome depicts an assortment of compositionally nonhomogeneous domains with numerous short compositionally homogeneous domains and relatively few long one (Elsik *et al*. 2009; Elhaik *et al*. 2010). Similar results were found for invertebrate genomes (Sodergren *et al*. 2006a; Sodergren *et al*. 2006b; Richards *et al*. 2008; Kirkness *et al*. 2010; Werren *et al*. 2010; Smith *et al*. 2011a; Smith *et al*. 2011b; Suen *et al*. 2011).

## 5. Applications of $GC_3$ in everyday's biology

Although the role of $GC_3$ as a proxy to large flanking regions was severely minimized, the question of which processes determine the GC content in 4-fold degenerate codons remains unanswered. It was therefore proposed to use $GC_3$ to detect sites under selection. The rationale is simple, if genomic GC-content is solely a consequence of mutation bias and the base composition is at equilibrium, then we expect equal numbers of synonymous mutations at 4-fold sites to be segregating within a species (Hildebrand, Meyer, and Eyre-Walker 2010), whereas deviation from such prediction may indicate selection. $GC_3$ remains a very useful tool to estimate codon usage bias and species diversity (Suzuki, Saito, and Tomita 2009).

$GC_3$ is useful for detection of recent horizontal gene transfer (HGT) events. Horizontal gene transfer occurs when an organism incorporates genetic material from another organism without being the offspring of that organism. Recently acquired genes retain nucleotide composition of the original genome and their identification is important for accurate reconstruction of phylogenetic trees, epidemiology, and genetic engineering.

$GC_3$ can also be used for gene prediction and genome annotation. In monocots, many genes demonstrate a negative GC gradient, that is, the GC content declines along the orientation of transcription. It is important to detect the presence of GC-rich sequences at the 5' end of genes because it influences the conformation of chromatin, the expression level of genes, and the recombination rate. Performance of genome annotation programs is often affected by 5'-3' gradients of nucleotide composition of coding region (Figure 1). Rare tissue-specific and stress-specific genes (that may not have mRNA support) are likely to belong to $GC_3$-rich class, and have a distinct pattern on the 5'-3'. If the gene-finding program is tailored to the more prevalent $GC_3$-poor genes, de-novo identification of these rare, but probably extremely important for stress response and adaptation, $GC_3$-rich genes will be hindered (Souvorov *et al* 2011).